\documentclass[%
reprint,
 amsmath,amssymb,
 aps,
]{revtex4-2}
\usepackage{CJK}
\usepackage{graphicx}% Include figure files
\usepackage{dcolumn}% Align table columns on decimal point
\usepackage{comment}
\usepackage{xcolor}
\usepackage[colorinlistoftodos]{todonotes}
\usepackage{hyperref}% add hypertext capabilities
\usepackage[english]{babel}
\usepackage{bm}% bold math
\usepackage{amsmath}
\usepackage{amssymb}
\usepackage{comment}
\DeclareUnicodeCharacter{03B2}{\ensuremath{\beta}}
\UseRawInputEncoding
\begin{document}

\preprint{APS/123-QED}

\title{Counterexample to the Bohigas Conjecture for Transmission Through a One-Dimensional Lattice}

\author{Ahmed A. Elkamshishy}
 \email{aelkamsh@purdue.edu}
 \affiliation{
 Department of Physics and Astronomy, Purdue University, West Lafayette, Indiana 47907 USA
}

\author{Chris H. Greene}%
 \email{chgreene@purdue.edu}
\affiliation{%
 Department of Physics and Astronomy, Purdue University, West Lafayette, Indiana 47907 USA
}%
\affiliation{Purdue Quantum Science and Engineering Institute, Purdue University, West Lafayette, Indiana 47907 USA}

\begin{abstract}
Resonances in particle transmission through a 1D finite lattice are studied in the presence of a finite number of impurities. Although this is a one-dimensional system that is classically integrable and has no chaos, studying the statistical properties of the spectrum such as the level spacing distribution and the spectral rigidity shows quantum chaos signatures. Using a dimensionless parameter that reflects the degree of state localization, we demonstrate how the transition from regularity to chaos is affected by state localization. The resonance positions are calculated using both the Wigner-Smith time-delay and a Siegert state method, which are in good agreement. Our results give evidence for the existence of quantum chaos in one dimension which is a counter-example to the Bohigas-Giannoni-Schmit conjecture.

\end{abstract}

\maketitle

In 1984, Bohigas, Giannoni and Schmit stated the celebrated (BGS) conjecture\cite{Bohigas_statistics} that describes the statistical properties of chaotic spectra. This conjecture draws a connection between quantum systems whose classical analog is chaotic and random matrix theory (RMT). Classical chaos is a consequence of the non-linearity of the Newtonian equations of motion, while Schr\"{o}dinger's equation is linear and strictly speaking has no chaos. Nevertheless, quantum signatures of chaos can arise and are exhibited by the statistical properties of the quantum energy level spectra, such as the level spacing distribution and the spectral rigidity (SR) \cite{SR_first_introduced,Bohigas_statistics}.

The spectrum of random matrices was first studied by Wigner in 1951 \cite{wigner-resonances,Meh2004,RMT}, who demonstrated the existence of a few universal classes based on the symmetry imposed on such matrices. In this paper, the two classes considered are the gaussian orthogonal ensemble (GOE) and the Poisson distribution.

The claimed connection between chaos and RMT is the essence of the BGS conjecture, which relates the spectral properties of quantum systems whose classical Hamiltonians are irregular (chaotic) to the GOE class. The BGS conjecture is formally stated as follows: Spectra of time-reversal-invariant systems whose classical analogs are K-systems\cite{ksystems} show the same fluctuation properties as predicted by GOE \cite{Bohigas_statistics}. On the other hand, the spectrum of a classically regular system follows Poissonian behavior \cite{Bohigas_statistics,chaos_and_quantum_thermalization,ChaosBook}.

The goal of the present study is to explore a one-dimensional quantum system whose classical analog is regular. Our analysis demonstrates that such systems have quantum chaos signatures. Moreover, two limiting cases are discussed: the {\it regular case} that agrees with the BGS conjecture, and a {\it second case} where a signature of chaos arises in the quantum spectrum, which matches the GOE statistics for both the nearest-neighbor distribution and the spectral rigidity. The chaotic behavior is shown to depend on a dimensionless parameter that represents the degree of state localization, thereby suggesting a connection between quantum chaos and the phenomenon of Anderson localization\cite{Anderson,Anderson_simple}. While RMT formally deals with discrete spectra only, our study discusses it in the context of very narrow resonances, which is justifiable because of their generally extreme narrowness.
%Due to the fact that RMT deals with discrete spectra, both bound and resonant states with bound character are included in our study.

The model considered in this study is a particle moving through a one-dimensional lattice with the lattice potential energy modeled by a sum of delta functions, one per lattice site. Thus the Hamiltonian is given by:

\begin{equation}\label{Hamiltonian}
    H = \frac{P^2}{2m} + {\sum_{n = -N/2}^{N/2}\alpha_n \delta(x-na)}.
\end{equation}

\noindent Here $N+1$ is the total number of lattice sites, and $\alpha_n$ is the strength of the $n$-th delta function. In the case of a perfectly clean periodic but finite lattice that we consider here, all $\alpha_n$ are the same and all are negative, but we will keep the notation general for now because part of our analysis will be an exploration of the effect of impurities. The reason behind choosing the delta function potential is the fact that an attractive delta function potential admits one bound state, so one can imagine the system as having one atomic state around each atomic site, which allows us to treat both the bound states and scattering dynamics. Modeling the lattice by considering one atomic state at each site and treating the effect of site-to-site tunneling as an effective hopping parameter has been studied for years and is referred to as the tight-binding approximation \cite{Anderson,Anderson_simple}. One of our goals is to compare the exact solution with the results of the tight-binding model. Our results show limitations of this approximation that become relevant in the context of transmission through a finite lattice.

Within the tight-binding approximation, two things are assumed: First, there exists one atomic state around each lattice site, and secondly, there is only hopping between nearest neighbors \cite{LCAO}.
The Hamiltonian in this case is written as: 
\begin{equation}\label{Anderson_Hamiltonian}
    H =\sum_i \epsilon_i a_i^\dagger a_i + \sum_{<i,j>} t_{i,j}a_i^\dagger a_j
\end{equation}

 \noindent where $\epsilon_i$ is the energy of the atomic state site $i$ and $t_{i,j}$ is the tunneling amplitude from site $j$ to site $i$, and the sum in the second term is taken for nearest-neighbors where $j = i \pm{1}$. Both can be calculated from the potential introduced in Eq.\ref{Hamiltonian}. If the lattice is periodic and all atomic sites are identical, then this model can be solved analytically \cite{AshcroftNeilW1976Ssp}. However, there is much interesting physics to study when impurities are placed in the lattice, such as the transport across the lattice. In that case, the periodicity is broken and there is no general analytical solution.  However, the spectrum can be obtained by writing the Hamiltonian in a matrix form and diagonalizing it \cite{Anderson_simple}.\\
 \indent The tight-binding approximation is useful in many cases, especially for the $N \rightarrow \infty$ limiting case. Most of the physics of bound states can be studied within this simple approximation. However, tight-binding has limitations, such as the fact that a set of $N$ negative delta functions do not necessarily support $N$ bound states. Even two delta functions in one dimension do not necessarily have two bound states, which can cause difficulties whenever finite lattices are considered. Secondly, within tight-binding, one can only study bound states. So, all the physics of scattering is missing, such as transmission resonances.
 
 The approach used in this paper is to consider a finite lattice and  treat it as a finite range potential. Then the scattering ($S$) matrix is obtained for a particle incident from $-\infty$ representing a scattering from the left, and for a particle incident from $+\infty$ representing scattering from the right, and all desired observables calculated. The resonance positions and widths can be calculated from either the Wigner-Smith time-delay \cite{wigner_time_delay,Q.delta,timedelay_in_one_d} or by imposing outgoing-wave Siegert state boundary conditions \cite{siegert,siegert_one_d}. A numerical solution is obtained for lattices with different values of the lattice size and the lattice constant. A main goal here is to study the real part of the resonance energy level distribution in the first energy band. In all calculations, atomic units are used, i.e. with $\hbar = a_0 = m_e =E_h= 1$.

The solution to the time-independent Schr\"{o}dinger equation for the Hamiltonian introduced in Eq.\ref{Hamiltonian} has the following form for particles incident from the left:
 \begin{equation}\label{wave_function}
     \psi(x) = \begin{cases} e^{iqx} + r(q)e^{-iqx} &\mbox{if } x  < \frac{-Na}{2}  \\
t(q)e^{iqx}  & \mbox{if } x  > \frac{Na}{2}  
\end{cases}
\end{equation}
where $r$ and $t$ are the reflection and transmission amplitudes respectively, and $q$ is the momentum of the incident particle. To obtain both $r$, and $t$, the solution inside the lattice has to be obtained. \\
\indent In the domain $x \in  [(n-\frac{1}{2})a,(n+\frac{1}{2})a]$,
\begin{equation}\label{soln_inside}
\Psi_n(x) = A_n e^{iqx} + B_n e^{-iqx} \equiv \psi_n^+ + \psi_n^-
\end{equation}
where 
$q = \frac{\sqrt{2mE}}{\hbar}$, and 
$m$ is the particle mass. After applying the the wave function continuity and derivative discontinuity conditions, the solutions in any two adjacent regions separated by a lattice constant $a$ are related by the transfer matrix \cite{transfer_matrix} as follows: 
\begin{equation}
\begin{split}
\begin{pmatrix}
    \psi_{n+1}^+ \\ \psi_{n+1}^-
    \end{pmatrix} & = \begin{pmatrix}
    e^{iqa} \frac{1+m\alpha_n}{iq} & \frac{m\alpha_n}{iq} \\  \frac{-m\alpha_n}{iq} &  e^{-iqa} \frac{1-m\alpha_n}{iq}
    \end{pmatrix}  \begin{pmatrix}
    \psi_{n}^+ \\ \psi_{n}^-
    \end{pmatrix} \\
     & \equiv T(\alpha_n,q) {\begin{pmatrix}
    \psi_{n}^+ \\ \psi_{n}^-  
    \end{pmatrix}}
\end{split}
\end{equation}
%The solution in Eq.\ref{wave_function} represents scattering by particles incident from the left. 
To obtain the full $S$-matrix, both transmission (reflection) amplitudes should also be obtained in the case of scattering by particles incident from the right, denoting them as $t'(q)$, and $r'(q)$. In one dimension, the $S$-matrix has the form \cite{S_Matrix_in_one_d,S_matrix}:  
 \begin{equation} \label{Smatrix}
     S = \begin{pmatrix}
    t & r' \\ r  & t'     
    \end{pmatrix}
 \end{equation}
\begin{figure}[h!]
%\begin{subfigure}{.5\textwidth}
  \includegraphics[width=1.0\linewidth]{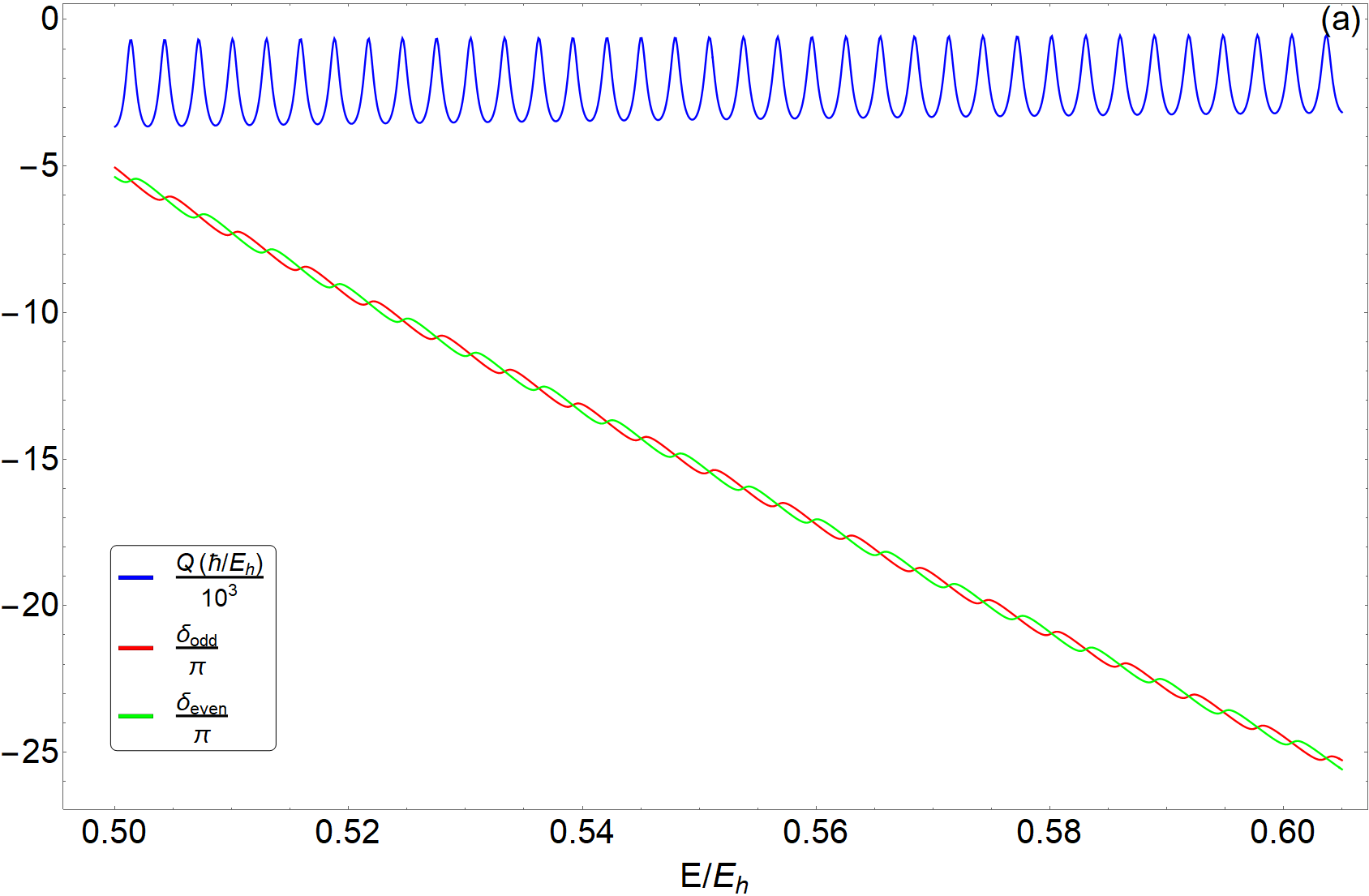} 
 % \caption{}
  %\label{fig:cleandelta}
%\end{subfigure}
%\begin{subfigure}{.5\textwidth}
  \includegraphics[width=1.0\linewidth]{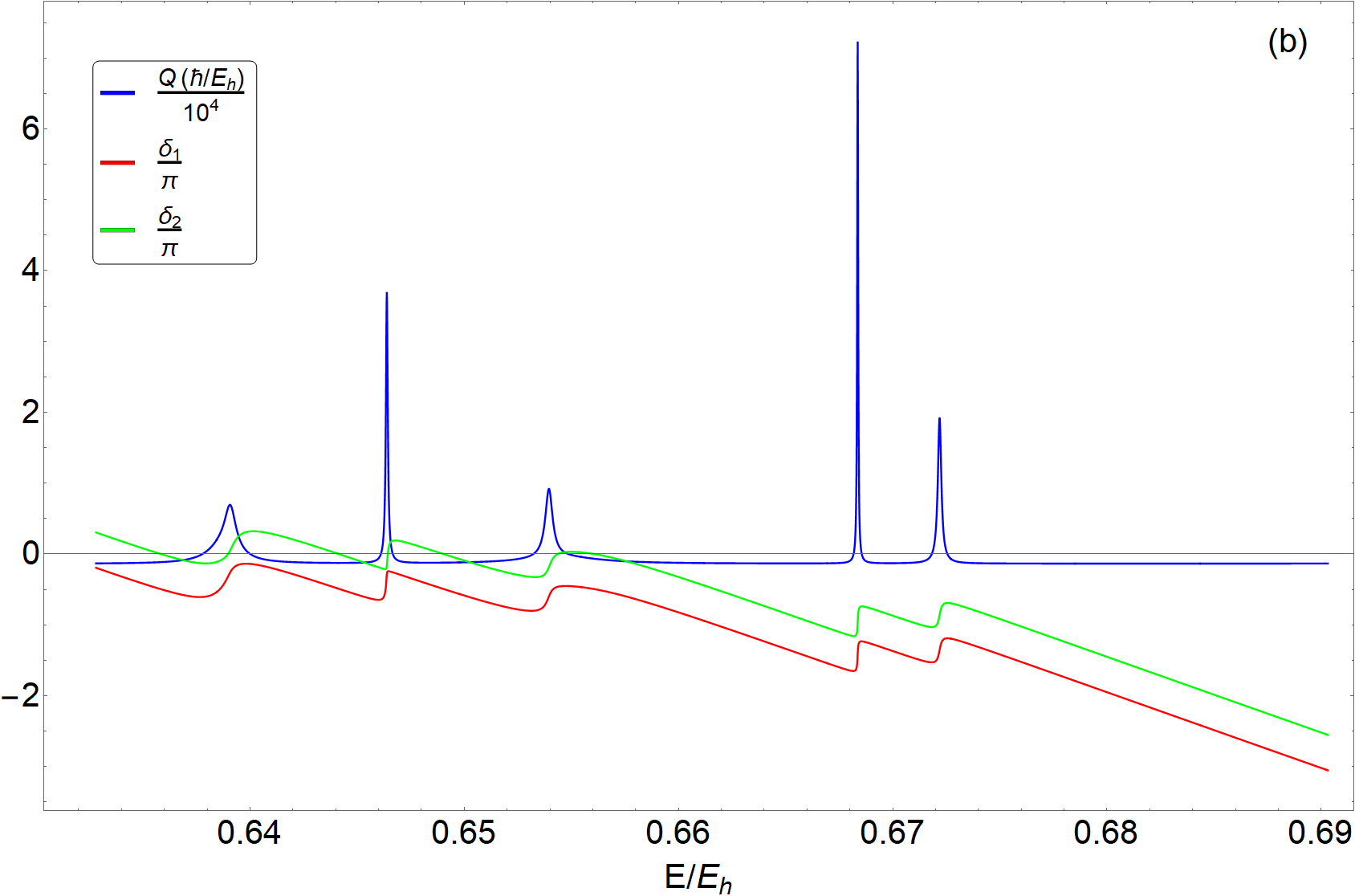}  
  \begin{frame}{}
  \includegraphics[width=1.0\linewidth]{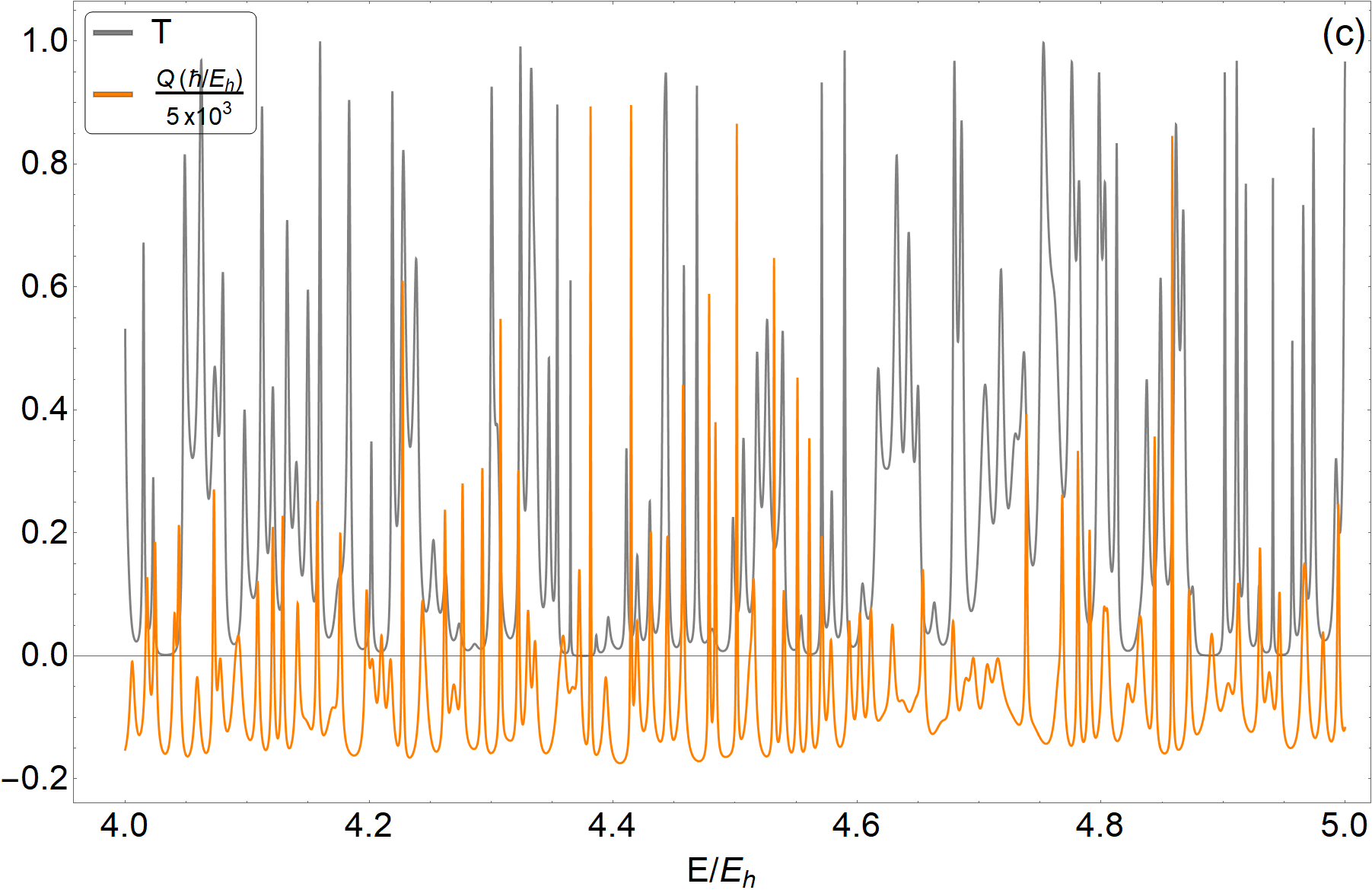}  
  \end{frame}
 % \caption{}
  %\label{fig:impuritiesDelta}
%\end{subfigure}
\caption{\label{fig:transmission} (a) The blue curve is the total time-delay, and the green (red) curves are the even (odd)-parity phase shifts for the periodic case with $\alpha_n = -1.5$ for all sites, $a = 0.8$, and $m = 1$. (b) The blue curve is the total time-delay, the red and the green curves are the eigenphase shifts for a lattice with impurities where $\alpha_n = -1.5 $ for $90 \%$ of the atoms and $\alpha_n = -1.9$ for the rest. The same mass and lattice constant as the periodic case are used. (c) The transmission probability and the total time-delay is plotted versus the collision energy where  $\alpha_n = -1.5 $ for $95 \%$ of the atoms and $\alpha_n = -1.9$ for the rest. The gray and the orange curves are the transmission coefficient and the total time-delay respectively.}
\end{figure}

 The distinctions between the periodic and the disordered cases become clear from plots of the Wigner-Smith time-delay \cite{wigner_time_delay}, $Q = i S \frac{dS^\dagger}{dE}$, and the phase shifts \cite{Q.delta}.The trace of $Q$ gives the total time-delay.  Fig.\ref{fig:transmission} shows that there are no resonances in the exactly periodic case with no impurities, and only a simple, regular oscillation of the total time-delay as a function of collision energy. On the contrary, when impurities are present, there are many narrow resonances. Associated with each resonance is a peak in the time-delay and a clear rise in the sum of the eigenphaseshifts  by $\pi$ radians as functions of energy. Studying the statistical properties of all the resonances in the first band shows chaos signatures in the system, as is demonstrated next.

The resonances can also be calculated using a different method: The Siegert state \cite{siegert_one_d} boundary conditions allows only outgoing waves, and they take the form:
 \begin{equation}\label{outgoing}
     \psi(x) = \begin{cases} Ae^{-iqx} &\mbox{if } x  < \frac{-Na}{2}  \\
Be^{iqx}  & \mbox{if } x  > \frac{Na}{2}  
\end{cases}
\end{equation}
With this boundary condition, the Hamiltonian is non-Hermitian and the spectrum is complex. Each eigen-value can be written as $ E_j ={E_0 }_j-i\Gamma_j /2$, where ${E_0}_j$ is the position of the $j^{\text{th}}$ resonance, and $\Gamma_j$ is the width \cite{moiseyev_2011,siegert_one_d}. After applying the boundary conditions in Eq.\ref{outgoing}, the energies are given by the roots of following equation:
\begin{equation}\label{determnetalcondition}
     M_{2,2} =0
\end{equation}
where $M$ is the total transfer-matrix given by $M(q) = \prod_{n=1}^N T(\alpha_n,q)$.\vspace{2pt}

Since the Hamiltonian in Eq.\ref{Hamiltonian} is one-dimensional and has no classical chaos \cite{ChaosBook}, the nearest neighbor distribution  of the resonances is expected to follow a Poisson distribution, $P(s) = e^{-s}$, and no level repulsion is expected \cite{Poisson_GOE}. On the other hand, classically chaotic systems are expected to have GOE statistics, and their level spacing distribution is expected to follow the Wigner-Dyson distribution \cite{Meh2004}, $P(s) = \frac{\pi}{2} s e^{-\frac{\pi s^2}{4}}$. The one key difference between the two distributions is that in chaotic systems, there is strong level repulsion \cite{Fano_Level_repulsion}, whereas this feature does not arise in classically regular systems.
Both distributions can be written more compactly in a convenient form as:
\begin{equation}\label{browdy}
     P^\beta (s)= b (1+\beta) s^{\beta} e^{-b s^{1+\beta}},\ \ {\rm with\ } b=\Gamma(\frac{2+\beta}{1+\beta})^{1+\beta},
\end{equation} 
%\begin{equation} \label{browdy2}
%     P^\beta (s)= (\Gamma(\frac{2 + \beta}{1 + \beta}))^{1 + \beta}} (1+\beta) s^{\beta} e^{-(\Gamma(\frac{2 + \beta}{1 + \beta})) s )^{1 +  \beta}}
% \end{equation}
 where $\beta = 0(1)$ corresponds to the Poisson (Wigner-Dyson) distribution. Equation \ref{browdy} is called the Brody distribution \cite{brody}. Following Wigner's early studies of RMT\cite{Meh2004,wigner-resonances}, extensive efforts have generalized the statistical properties of both Hermitian and non-Hermitian systems \cite{RMT,non-hirmetian-RM,complex-nonlinear,bohigas-fluctuations,Characteristics_of_level_spacing_statistics_in_chaotic_graphene_billiards,closedorbits.74.1538}.
 
\begin{figure}[h!]
%\begin{subfigure}{.5\textwidth}
  \includegraphics[width=1.0\linewidth]{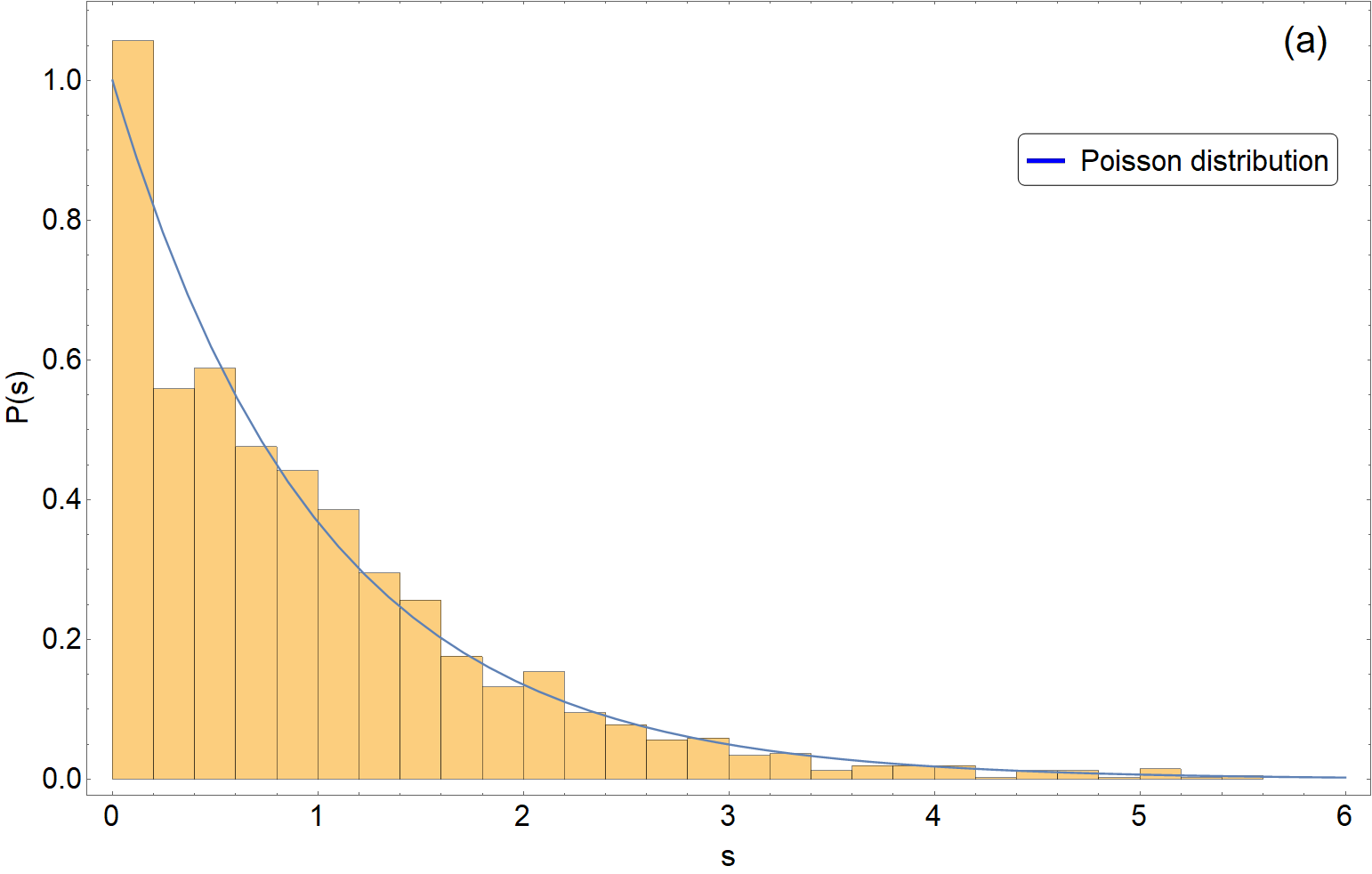} 
 % \caption{}
  %\label{fig:poisson}
%\end{subfigure}
%\begin{subfigure}{.5\textwidth}
  \includegraphics[width=1.0\linewidth]{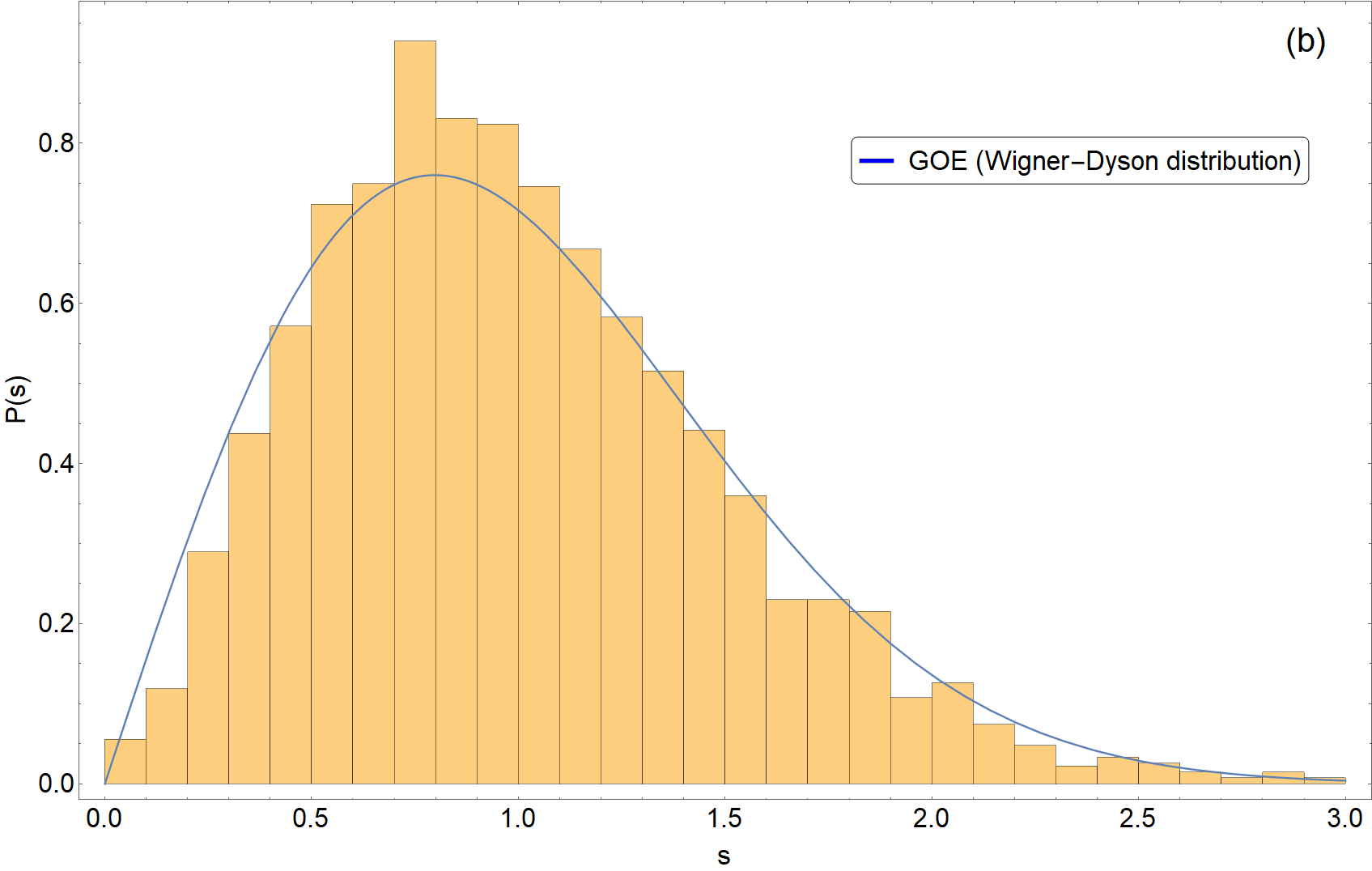}  
 % \caption{}
  %\label{fig:wigner}
%\end{subfigure}
%\begin{subfigure}{.5\textwidth}
  \includegraphics[width=1.0\linewidth]{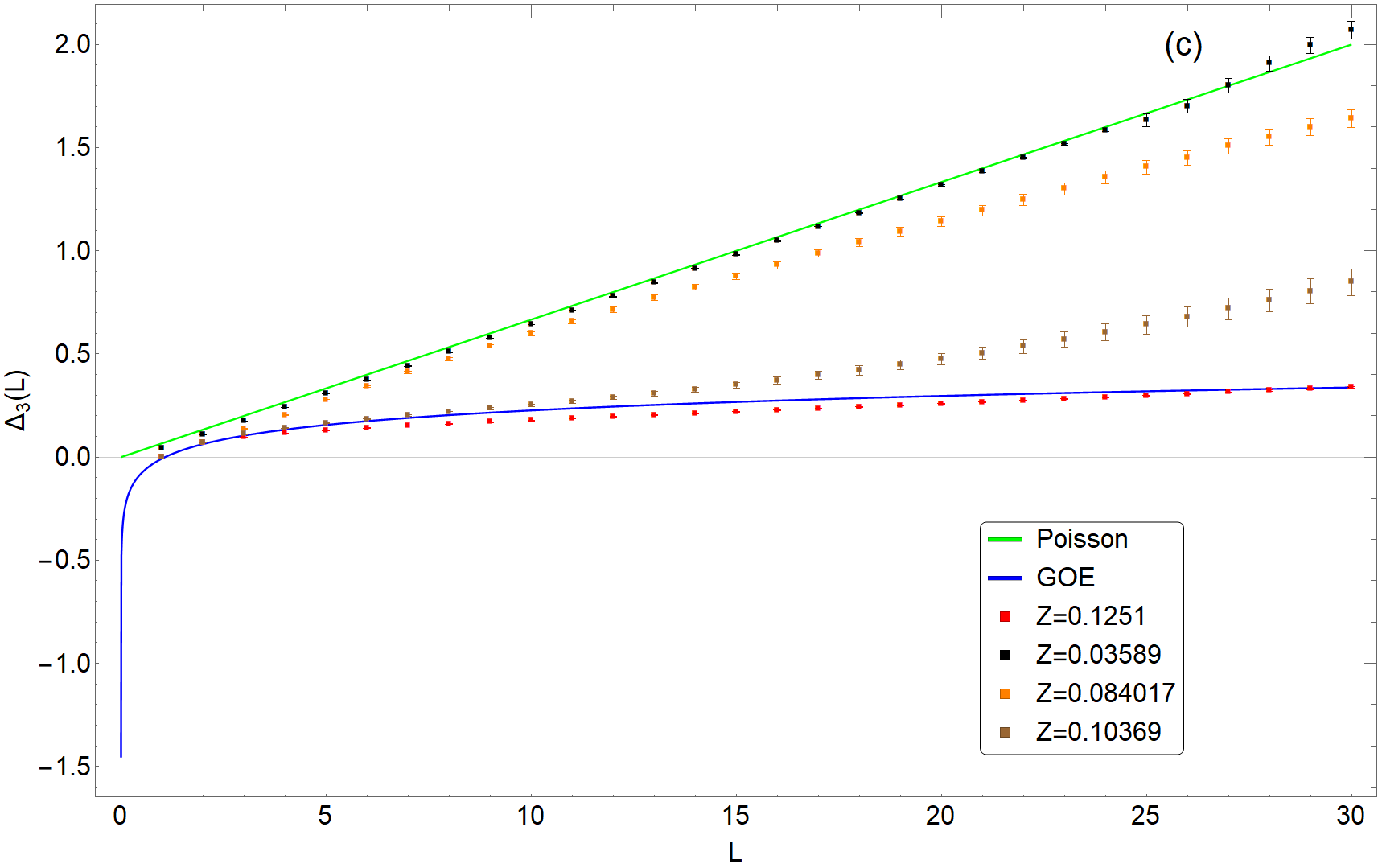} 
 % \caption{}
  %\
%\end{subfigure}
\caption{\label{fig:SR} (a) The nearest-neighbor spacing distribution of the resonances is shown for $Z = 0.0358$.  The blue curve is the Brody distubution for $\beta = 0$ (Poisson). The bars show the calculated nearest-neighbor level distribution obtained from the solution of Eq.\ref{determnetalcondition}. (b) The level spacing distribution of the resonances is shown for $Z= 0.1251$. The blue curve is the Brody distribution for $\beta = 1$ (Wigner-Dyson). The bars show the calculated nearest-neighbor level distribution obtained from the solution of Eq.\ref{determnetalcondition}. (c) The spectral rigidity for different values of $\beta$. The red dots are the values of the SR calculated for $Z = 0.1251$ matching the GOE curve, while the black dots are those calculated for $Z = 0.0358$ matching the Poisson curve}
\label{fig:dist}
\end{figure}

  The results obtained from both the time-delay analysis and Eq.\ref{determnetalcondition} agree, and they show that the nearest neighbor spacing (NNS) distribution depends strongly on the extent to which the energy eigenstates are localized. For different lattice parameters, two limiting cases emerge, corresponding to $\beta = 0$ or $1$. To see the dependence quantitatively, it is convenient to define the dimensionless parameter $Z=\frac{<\sigma_x>}{L}$ , where $<\sigma_x>$ is the average uncertainty in the position, taken over the domain $x \in  [-\frac{L}{2},+\frac{L}{2}]$, averaged over all resonance states in the first band, and $L = Na$ is the length of the lattice. Evidently, $Z$ represents a statistical measure of how localized are the resonance energy eigenstates. Table \ref{tab:Table1} shows that the transition between regularity and chaos occurs in a way that is consistent with the claim that the quantum chaos signatures depend on the state localization. Moreover, some intermediate values of $\beta$ from fitting the NNS distribution are also obtained for different lattices with different values of $Z$.
 In addition to the nearest-neighbor spacing distribution, the calculation of the spectral rigidity \cite{SR_first_introduced,Berry_Semi_Classical} shows similarly good agreement for the two limiting cases, and each one corresponds to the same statistics consistent with the nearest-neighbour distribution as shown in Fig.\ref{fig:dist}.The results, shown in both Fig.\ref{fig:SR}, and Table.\ref{tab:Table1}, are calculated only for resonance states while bound states (which are classically forbidden in this system) are not considered. Hence, our claim of quantum chaos signatures for this classically nonchaotic system applies only to resonance states because only positive energy solutions have a nontrivial classical analogue. 

\begin{table}[ht]
\caption{The transition, as measured by $\beta$ in Eq.\ref{browdy}, from the Poisson to Wigner distribution is presented as a function of the localization parameter $Z=\frac{<\sigma_x>}{L}$. The values in this table are taken for different numbers of impurities, namely from $5\%$ to $20\%$.} 
\centering
\label{tab:Table1}
\resizebox{1\columnwidth}{!}{\begin{tabular}{l | l | l | l | l | l | l| l| l| l}
$Z$ & 0.0298 & 0.0358 & 0.0483 & 0.0840 & 0.0987 & 0.1040 & 0.1105 & 0.1239 & 0.1251 \\
\hline
$\beta$ & 0 & 0 & 0.2 & 0.4 & 0.5  & 0.7 & 0.9 & 1 & 1 \\ \hline
$\Delta \beta$ & 0.008 & 0.012  & 0.029 & 0.020 & 0.060 & 0.011 & 0.014 & 0.013 & 0.007  \\
\hline
\end{tabular}}
\end{table}
\noindent The fractional values of $\beta$ do not correspond to any of the RMT classes \cite{RMT}, however, the values help to visualize how chaos emerges in the system.

Classical chaos is absent in one dimension because of the small number of degrees of freedom, which implies that any small change in the initial condition cannot produce a drastic change in the classical trajectory of the particle. In other words, the Lyapunov exponent \cite{ChaosBook,Quantum_chaos_gutzwiller,chaos_and_quantum_thermalization,phasespaceQM,brack:semiclassical} always vanishes in any systems whose classical Hamiltonian is given by Eq.\ref{Hamiltonian} with the replacement of each delta function by a very narrow Gaussian or any other attractive well. The meaning of the results in both Table \ref{tab:Table1} and Fig.\ref{fig:SR} is that there exists a one-dimensional system that is classically regular but which displays the signatures of quantum chaos. This surprising result serves as a powerful counter example to the BGS conjecture in this remarkably simple one-dimensional system. Examples of so-called ``quantum chaos" have been studied in detail in more complex systems such as a Rydberg atom in a magnetic field, three-dimensional lattices, and chaotic systems exhibiting closed orbit signatures\cite{hydrogen_in_magnetic_field,chaos_in_three_d_lattices,onset_of_quantum_chaoes_in_one_d_lattice,GOE_in_one_d}. But those systems mentioned are either higher-dimensional or else many body systems whose classical analogs are irregular. A recent study by Ujfalusi and Varga \cite{chaos_in_one_d?} explores whether there is any one-dimensional systems whose statistical spectrum exhibits chaotic signatures. In that paper a Wigner-Dyson distribution for the level spacing of the energies was assumed and then the potential energy curve was derived, obtaining a resulting potential curve with many sharp peaks.  Their result is consistent with the results in the present study, in particular for Hamiltonians with sharp irregular shapes like the delta function. One can view the treatment  in \cite{chaos_in_one_d?} as essentially solving the inverse problem of quantum chaos in one-dimension.
  \begin{figure}[h! ]
\begin{frame}{}
  \includegraphics[width=1.0\linewidth]{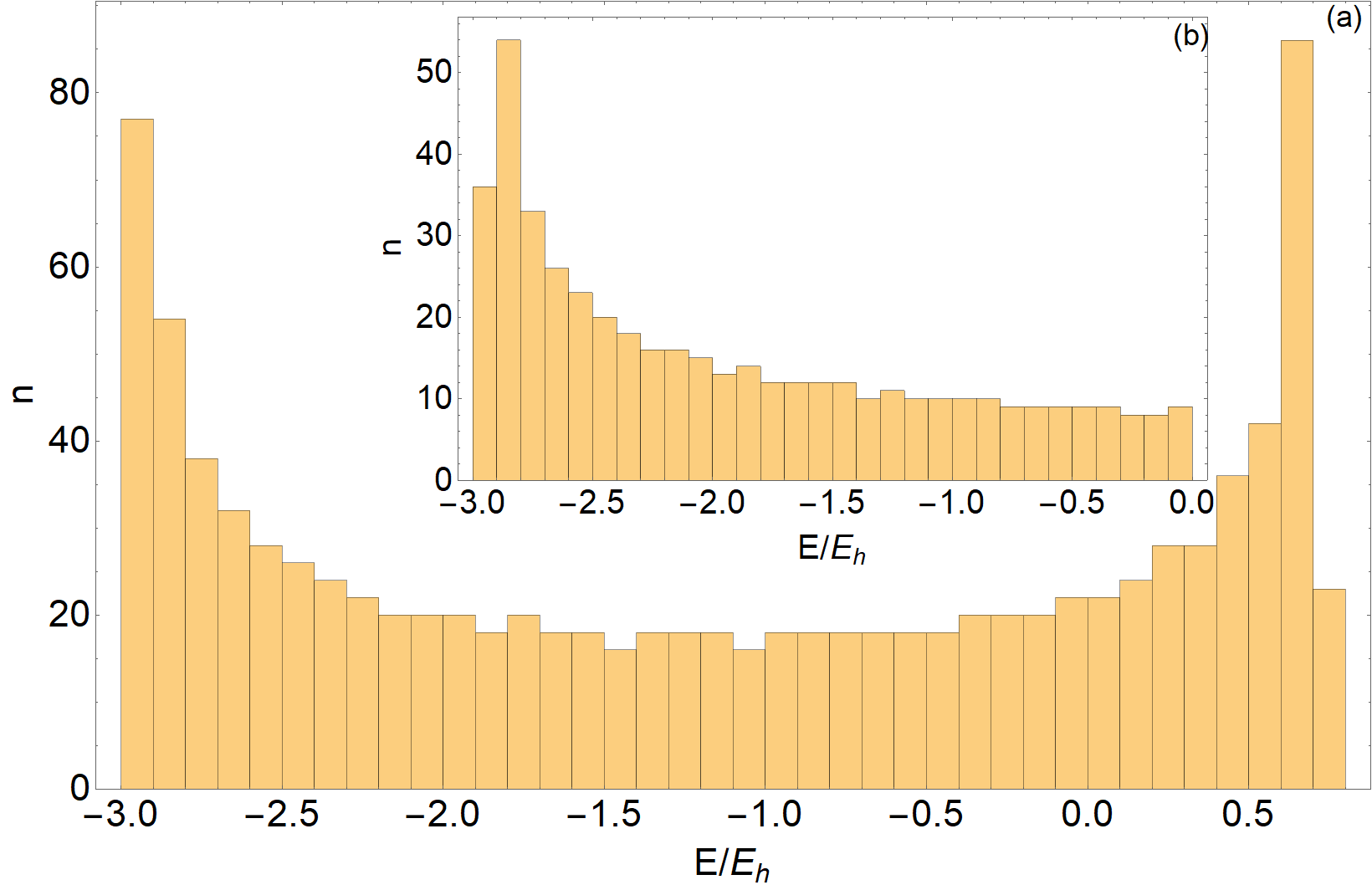} 
  \end{frame}
%\begin{frame}{}
 % \includegraphics[width=1.0\linewidth]{DOS exact solution.png}  
  %\end{frame}
\begin{frame}{}
   \includegraphics[width=1.0\linewidth]{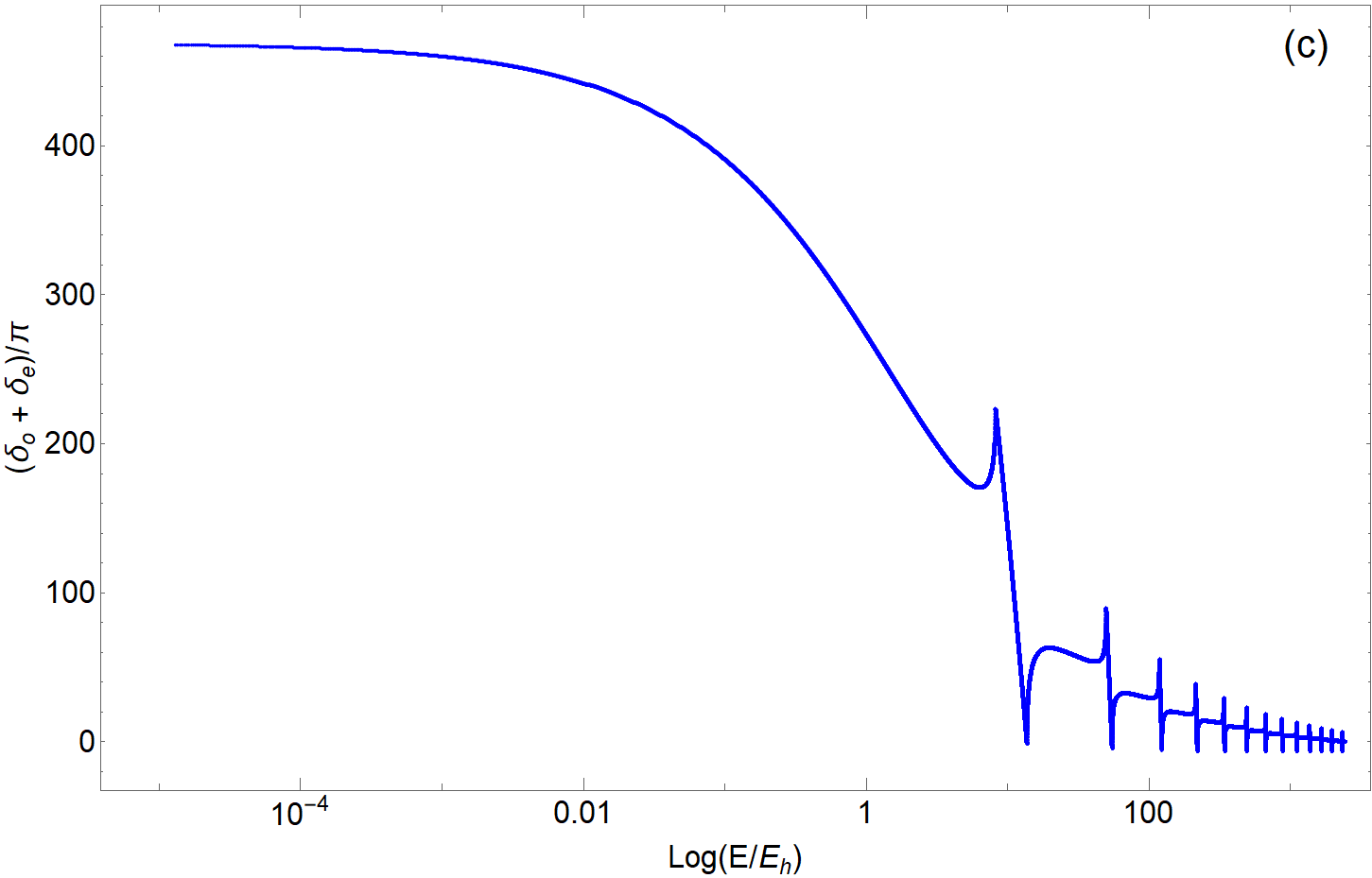}
   \end{frame}
\caption{\label{fig:DOS} (a) The density of bound states plotted versus the energy. The bars show the number of states within an energy interval, calculated from the tight-binding approximation. (b) The density of bound states plotted versus the energy. The bars show the number of states within an energy interval, obtained from the poles of the $S$-matrix in the complex energy plane. (c) The sum of the even and odd phase shifts versus the logarithm of the energy. The value of the sum of the phase shifts at zero energy fixes the number of bound states and gives the same number of bound states predicted by Levinson's theorem, namely 463 states.} 
\end{figure}

In one-dimensional infinite lattices, all states are localized in the presence of {\it any} percentage of impurities as shown by Anderson \cite{Anderson,Anderson_simple}. Anderson localization has been studied in many one-dimensional systems \cite{Anderson_simple,timedelay_transport}, and in most of the cases studied in the literature, the tight-binding approximation is implemented with either periodic or vanishing boundary conditions. Many results in the literature document the lack of diffusion in one dimension. However, in Fig.\ref{fig:transmission}, the transmission probability $T = \cos^2(\delta_1 - \delta_2)$ obtained from the eigenphase shifts and determined by our choice of the channel functions, is enhanced and approaches unity for the narrow resonances. The study of how disorder affects the transport has been explored in details, in terms of quantities like the Wigner-Smith time-delay and the Thouless conductivity \cite{Thouless_1972,THOULESS197493,timedelay_distribution_in_oned,timedelay_transport}, deriving there a relation between the localization length and the disorder. However, it has usually been assumed that the disorder in the system is taken from a random distribution, and all the quantities of interest such as the conductivity or the localization length are derived based on that assumption. As was mentioned previously, the strength of the impurities is not taken randomly in our study; only two different kinds of atoms have been assumed to be present, and the strength of all the impurities has been taken to be the same, while those impurities are placed randomly throughout the lattice. This is why the dimensionless quantity $Z$ introduced in Table \ref{tab:Table1} is - more conveniently - chosen as our measure of the localization.

 The scattering solutions obtained from the $S$-matrix analysis can also be used to calculate bound states by searching for the poles of $S(E)$ in the complex energy plane, and comparing them with the spectrum obtained from the tight-binding approximation. The Hamiltonian in Eq.\ref{Anderson_Hamiltonian} admits a number of bound states that is always equal to the number of lattice points. Stated differently, the Hilbert space of the particle on the lattice is given by the direct sum of every single particle Hilbert space around each lattice point\cite{AshcroftNeilW1976Ssp}. Consequently, if each attractive delta function admits one bound state, then the prediction of the tight binding gives a number of bound states that is equal to the number of lattice points. Fig.\ref{fig:DOS}, shows the difference between the density of states of the bound states between the solution obtained from the $S$-matrix and the tight-binding model. Both are calculated for a periodic lattice with $N = 1000$. The main difference is in the number of bound states. As argued above, the tight-binding gives $1000$ bound states. However, there are only $463$ bound states obtained from the $S$-matrix treatment, while the rest of eigenstates with bound character resonances and are only quasi-bound. Moreover, as is shown in Fig.\ref{fig:DOS}, the number of bound states can be determined by the value of the eigenphase shifts at zero energy, as is predicted by Levinson's theorem in one dimension \cite{Levinson_one_d}. After setting the values of both phase shifts at infinite energy to zero, we have: \begin{equation}\label{Levinson's theorem}
\frac{\delta_o (0)  + \delta_e (0)}{\pi} =(N_b + \frac{1}{2}) = (463+\frac{1}{2})   
\end{equation} 
where $N_b$ is the total number of bound states and that gives exactly the same number of bound states; This Levinson's theorem result is confirmed in our study. The purpose of the above discussion is to demonstrate the internal consistency of the method that has been used to obtain all of the results shown in this paper. 

In conclusion, the results of the calculations and the arguments mentioned above are a clear counterexample to the Bohigas conjecture, as they show that quantum chaos signatures do appear in a one-dimensional system that has no classical chaos. Moreover, the results show the additional physics that can be examined by going beyond the tight-binding approximation, particularly if the system of interest is a lattice of finite size. Anderson's work \cite{Anderson} in the 1960s demonstrated that in one-dimensional disordered lattices there is no transport even for the smallest amount of any disorder that breaks the periodic symmetry of the lattice. Anderson proved his statement mathematically by considering only bound states that form a conduction band in the clean case. As is shown in Table \ref{tab:Table1}, however, the signatures of chaos in this system hinge critically on the quantitative extent of state localization. These results show the value of considering aspects of such systems that go beyond the tight-binding approximation, such as studying the transmission resonances that are the focus of the present exploration. 

Informative discussions with Francis Robicheaux are greatly appreciated. This work has been supported by NSF grant No.PHY-1912350.

\bibliographystyle{unsrt} % Use for unsorted references  
%\bibliographystyle{plainnat} % use this to have URLs listed in References
%\cleardoublepage
\bibliography{bibliography} % Path to your References.bib file
\nocite{*}

\end{document}